\documentclass[aps,prl,twocolumn,showpacs]{revtex4-1}
\usepackage{graphicx}
\usepackage{amsmath,amsfonts}

\begin{document}

\title{Exciton supersolidity in hybrid Bose-Fermi systems}

\author{Micha\l {} Matuszewski}
\affiliation{Instytut Fizyki PAN, Aleja Lotnik\'ow 32/46, 02-668 Warsaw, Poland}

\author{Thomas Taylor}
\affiliation{School of Physics and Astronomy, University of Southampton, Southampton, SO171BJ, United Kingdom}

\author{Alexey V. Kavokin}
\affiliation{School of Physics and Astronomy, University of Southampton, Southampton, SO171BJ, United Kingdom}

\begin{abstract}
We investigate the ground states of a Bose-Einstein condensate of indirect
excitons coupled to an electron gas. We show that in a properly designed
system, the crossing of a roton minimum into the negative energy domain can
result in the appearance of the supersolid phase, characterized by
periodicity in both real and reciprocal space. Accounting for the
spin-dependent exchange interaction of excitons we obtain ferromagnetic
supersolid domains. The Fourier spectra of excitations of weakly perturbed
supersolids show pronounced diffraction maxima which may be detected
experimentally.
\end{abstract}

\maketitle

\begin{figure}[tbp]
\includegraphics[width=8.5cm]{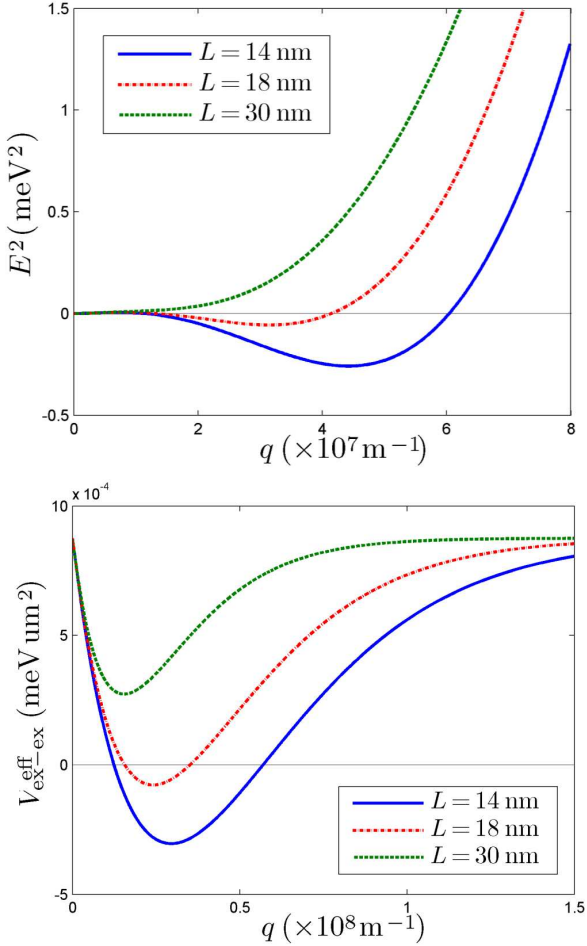}
\caption{ The excitation spectrum (top) and effective exciton-exciton
interaction potential (bottom) for three values of the exciton-electron
layer separation $L$. We use the exciton mass $m_{\mathrm{ex}} = 0.21 m_{%
\mathrm{e}}$, and the average exciton density $\overline{n}_{\mathrm{tot}} =
10^{11} \mathrm{cm}^{-2}$. The exciton dipole separation, $l=12 \mathrm{nm}$%
, and the correction factor, $\protect\xi=0.05$.}
\label{potentials}
\end{figure}

A supersolid phase is characterised by superfludity and long range order in
both real and reciprocal spaces \cite{Supersolid_old}. One of the suggested
indications of this phase transition is a "rotonlike instability" observed
if the energy of the roton minimum of a superfluid is pushed below the
energy of the ground state (characterised by zero wave-vector, $q$) \cite%
{Rotonlike}. Strong indications of the phase tranistion to supersolidity
have been found in $^{4}$He at low temperatures and high pressure \cite%
{Supersolidity_He}. Until now, experimental studies of supersolidity have
been limited to $^{4}$He and $^{3}$He, while, potentially, this phase may
occur in any system of interacting bosons \cite{Lozovik2011}. Here we
discuss the possibility of formation of exciton supersolids in specially
designed semiconductor structures.

Recently, Bose-Einstein condensation (BEC) and superfluidity of crystal
quasiparticle excitations - excitons or exciton polaritons - have been
reported in various semiconductor systems \cite{Butov, Kasprzak, Kavokin,
Amo}. The exciton-exciton interaction is usually almost independent of the
wave vector \cite{Ciuti}, which is probably why in most experimentally
studied cases the roton minimum is not formed. Nevertheless, formation of a
roton minimum and even of a roton instability has been predicted for a
hybrid system where a two-dimensional (2D) exciton condensate is placed in
close vicinity to a 2D electron gas \cite{Rotons_Hybrid}. Moreover, the
recent theoretical paper by Parish \textit{et al} \cite{Parish} evokes the
possibility of the supersolid phase for excitons in asymetrically populated
bilayers. Here we show that an exciton condensate in a hybrid system where
excitons and an electron gas interact but do not overlap can undergo a
transition to the supersolid phase.

We shall consider a system of coupled semiconductor quantum wells (CQW)
which contain the BEC of spatially indirect excitons which has been widely
studied in the past \cite{Lozovik}. We introduce an important new element to
this system: a 2D gas of  free electrons located in a thin metallic layer
grown on the top of semiconductor structure. The metal is separated from CQW
by a wide band gap semiconductor layer, so that there is no overlap between
free carriers and excitons, while they are close enough to each other to
allow for efficient Coulomb interaction. The system is biased in such a way
that the exciton dipole moments are oriented towards metal. This
configuration prevents attraction between excitons and electrons and
formation of trions. We phenomenologically describe spin-dependent
interactions of dipole excitons by two constants, as specified below, and
concentrate on the new physics brought by exciton-electron interactions. The
system we study has been proposed for realization of charged electron-hole
complexes \cite{Yudson} and of superconductivity mediated by the exciton
condensate \cite{Laussy}. Compared to the unequally populated electron-hole
bilayers considered by Parish \textit{et al} \cite{Parish} this system has
an evident advantage of protecting excitons from the phase space filling
effect, which would allow operation with much higher exciton and 2D electron
gas densities than in any system where excitons and free carriers spatially
overlap. In the present work we concentrate on the phase trasitions of the
exciton condensate induced by the presence of an electron gas in the
neighboring metallic layer. Referring to the most popular GaAs based system,
we account for the exchange interaction between excitons having $+1$ and $-1$
spin projections to the structure axis. This interaction, which represents
the superexchange via intermediate dark excitons with spin projections $+2$
and $-2$, has been recently addressed theoretically \cite{Combescot,
Sinclaire}.

An exciton condensate is described by a factorized Hartree-Fock ansatz $\Psi
_{\pm }(\mathbf{r}_{1},\dots ,\mathbf{r}_{N},t)=\prod_{i}\psi _{\pm }(%
\mathbf{r}_{i},t)$~\cite{Leggett_BEC}, which, in the limit of large number
of particles $N$, leads to the following mean field energy with a nonlocal
interaction potential, where $\pm $ denotes the $+1$ and $-1$ exciton spin
states, 
\begin{align}
& H_{\pm } =N_{\pm }\int d\mathbf{r}\,\psi _{\pm }^{\ast }(\mathbf{r}%
,t)\left( -\frac{\hbar ^{2}}{2m_{\mathrm{ex}}}\nabla ^{2}+\right.  \label{H}
\\
& +\left. \frac{1}{2}\int d\mathbf{r}^{\prime }\left( V_{\mathrm{ex-ex}}^{%
\mathrm{eff}}(\mathbf{r}-\mathbf{r}^{\prime })n_{\mathrm{tot}}(\mathbf{r}%
^{\prime },t)+\alpha n_{\mp }(\mathbf{r^{\prime }},t)\right) \right) \psi
_{\pm }(\mathbf{r},t).  \notag
\end{align}
$\alpha $ denotes the interaction constant between the two exciton spin
components, $n_{\mathrm{tot}}(\mathbf{r},t)=n_{\mathrm{+}}(\mathbf{r},t)+n_{%
\mathrm{-}}(\mathbf{r},t)=N_{+}|\psi _{\mathrm{+}}(\mathbf{r}%
,t)|^{2}+N_{-}|\psi _{\mathrm{-}}(\mathbf{r},t)|^{2}$, and $V_{\mathrm{ex-ex}%
}^{\mathrm{eff}}(\mathbf{r})$ is the spin-isotropic exciton-exciton
interaction potential. The corresponding time-dependent Gross-Pitaevskii
(GP) equations for the two condensates read 
\begin{align}
& i\hbar \frac{\partial \psi _{\pm }(\mathbf{r},t)}{\partial t}=-\frac{\hbar
^{2}\nabla ^{2}}{2m_{\mathrm{ex}}}\psi _{\pm }(\mathbf{r},t)\,+  \label{GP}
\\
& +\int V_{\mathrm{ex-ex}}^{\mathrm{eff}}(\mathbf{r}-\mathbf{r}^{\prime })n_{%
\mathrm{tot}}(\mathbf{r}^{\prime },t)d\mathbf{r}^{\prime }\psi _{\pm }(%
\mathbf{r},t)+\alpha n_{\mp }(\mathbf{r},t)\psi _{\pm }(\mathbf{r},t)\,. 
\notag
\end{align}

Derivation of the effective exciton-exciton interaction potential in this
system has been recently published \cite{Rotons_Hybrid}. The matrix element
of effective exciton-exciton interaction depending on frequency and
wave-vector reads 
\begin{align}
& V_{\mathrm{ex-ex}}^{\mathrm{eff}}(q,\omega )  \label{potential} \\
& =\frac{\left( V_{22}+\frac{V_{12}^{2}(q)\Pi _{1}(q,\omega )}{%
1-V_{11}(q)\Pi _{1}(q,\omega )}\right) \left( (\hbar \omega
)^{2}-(E^{ex}(q))^{2}\right) }{(\hbar \omega )^{2}-(E^{ex}(q))^{2}-2N_{0}%
\left[ V_{22}+\frac{V_{12}^{2}(q)\Pi _{1}(q,\omega )}{1-V_{11}(q)\Pi
_{1}(q,\omega )}\right] E^{ex}(q)},  \notag
\end{align}%
where 
\begin{equation*}
\Pi _{1}(\mathbf{q},\omega )=\sum_{\mathbf{k}}\frac{f_{\mathbf{k-q}}-f_{%
\mathbf{k}}}{\hbar \omega +i\hbar \delta +E_{\mathbf{k-q}}^{el}-E_{\mathbf{k}%
}^{el}},
\end{equation*}%
\begin{equation}
V_{11}=\frac{e^{2}}{2\epsilon _{0}\epsilon A}\cdot \frac{1}{q}
\end{equation}%
with $\epsilon $ denoting the dielectric constant of the media and $A$ the
sample area. The matrix element of exciton-exciton interaction $V_{22}$ is
dominated by dipole-dipole repulsion \cite{Laikhtman,Zimmermann} and can be
expressed as%
\begin{equation}
V_{22}(q)=\frac{ed\xi }{\epsilon _{0}\epsilon A},
\end{equation}%
where $d$ is the exciton dipole moment in normal to the QW plane direction, $%
\xi $ accounts for the reduction of dipole-dipole repulsion due to
electron-hole pair correlations in the exciton condensate \cite{Zimmermann}.
The expression for the matrix element of electron-exciton interaction reads
(the details of the calculation can be found in e.g. Ref.\onlinecite{Ramon}%
): 
\begin{align}
V_{12}& (q)= \\
& \frac{ede^{-qL}}{2\epsilon _{0}\epsilon A}\left\{ \frac{\beta _{e}}{\left[
1+(\frac{\beta _{e}qa_{B}}{2})^{2}\right] ^{3/2}}+\frac{\beta _{h}}{\left[
1+(\frac{\beta _{h}qa_{B}}{2})^{2}\right] ^{3/2}}\right\} +  \notag \\
& \frac{e^{2}e^{-qL}}{2\epsilon _{0}\epsilon qA}\left\{ \frac{1}{\left[ 1+(%
\frac{\beta _{e}qa_{B}}{2})^{2}\right] ^{3/2}}-\frac{1}{\left[ 1+(\frac{%
\beta _{h}qa_{B}}{2})^{2}\right] ^{3/2}}\right\}   \notag
\end{align}%
where $\beta _{e,h}=m_{e,h}/(m_{e}+m_{h}),$ with $m_{e,h}$ being the
effective masses of electrons and holes.

The poles of the effective potential determine the dispersions of the
collective modes of the exciton system given by the equation 
\begin{equation}
(\hbar \omega )^{2}=(E^{ex}(q))^{2}+2N_{0}\tilde{V}_{22}\left( q\right)
E^{ex}(q),  \label{disp}
\end{equation}
where 
\begin{equation}
\tilde{V}_{22}\left( q\right) =V_{22}+V_{12}^{2}(q)\Pi _{1}(q,\omega
)/[1-V_{11}(q)\Pi _{1}(q,\omega )].  \label{effpot}
\end{equation}

\bigskip In general, the exciton-exciton interaction mediated by the
electron gas is a time-dependent, retarded interaction. In order to obtain a
non-retarded GP equation (\ref{GP}), we average the potential (\ref%
{potential}) over a frequency range limited by the exciton dissociation
frequency $\omega _{d}=E_{B}/\hbar $. Depending on the frequency-averaged
interaction potential $V_{\mathrm{ex-ex}}^{\mathrm{eff}}(\mathbf{q})$, the
excitation spectrum of the condensate coupled to an electron gas can develop
a roton minimum due to the attractive character of the electron mediated
interactions~\cite{Rotons_Hybrid}, see Fig.~\ref{potentials}. This Figure
shows the exciton dispersion curves calculated from Eq. (\ref{disp}) in
comparison with the average interaction potential for different separations $%
L$ between the QWs containing the 2D electron gas and the exciton
condensate. Indeed, for small\ separation between excitons and electrons the
potential becomes attractive in an intermediate range of $q$ and a roton
instability develops.

\begin{figure}[tbp]
\includegraphics[width=8.5cm]{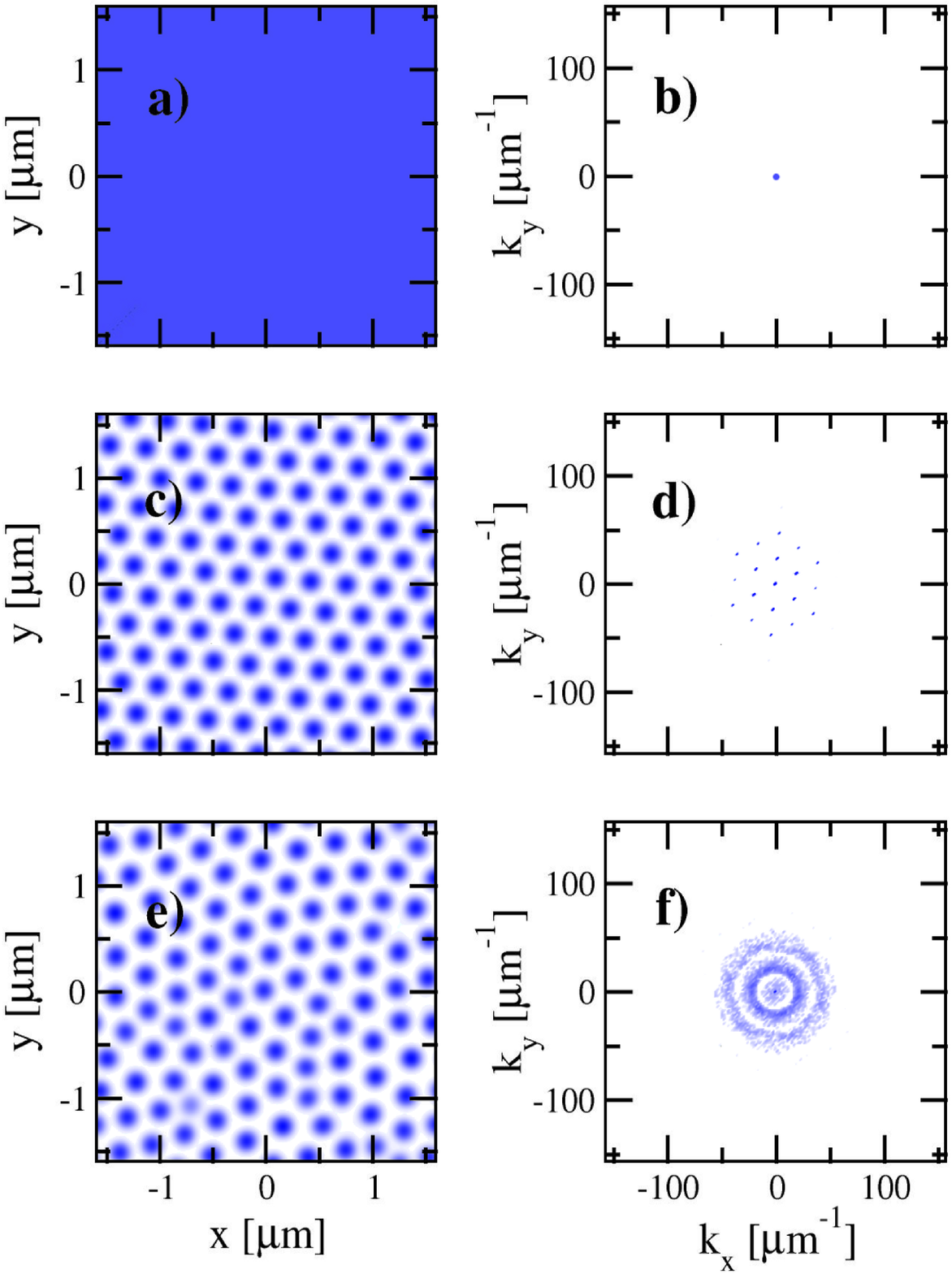}
\caption{Examples of density profiles of the exciton condensate in real
space (left column) and reciprocal space (right column, density in
logarithmic scale). The figures (a), (b) show the homogeneous ground state
for separation $L=30\,$nm, and the figures (c), (d) depict the supersolid
ground state for $L=18\,$nm, characterized by periodicity in both real and
momentum spaces. The orientation of the supersolid lattice is chosen
spontaneously. (e), (f) A metastable state with lattice imperfections,
corresponding to a local energy minimum. In momentum space, such a state is
characterized by concentric rings.}
\label{profiles}
\end{figure}

The existence of a roton instability has been connected with the possible
existence of a supersolid phase~\cite{Supersolid_old,Rotonlike}, in which
superfluid and crystalline orders coexist. We investigate this possibility
by numerical minimization of the Hamiltonian~(\ref{H}) using the imaginary
time propagation method. To start with we assume $\alpha =0$ and investigate
a single-component exciton condensate.\ We determine the ground states for
various physical parameters in the case of a two dimensional box with
periodic boundary conditions. We take realistic values of parameters $m_{%
\mathrm{ex}}=0.21m_{\mathrm{e}}$, $m_{\mathrm{e}}$ being the free electron
mass, the average exciton density $\overline{n}_{\mathrm{tot}}= 10^{11}%
\mathrm{cm}^{-2}$, with an exciton dipole separation $l=12\,$nm, and
different effective interaction potentials corresponding to various
exciton-electron layer separation $L$.

We distinguish three regimes in which the minimization procedure leads to
qualitatively different results. In the case of a positive roton gap, when
the excitation spectrum does not cross into the negative energy domain, the
ground state is a homogeneous superfluid, as depicted in Figs.~\ref{profiles}%
(a) and (b) for $L=30\,$nm. When decreasing the interlayer distance to $%
L=18\,$nm, a dramatic change occurs, and a state characterized by a periodic
triangular lattice becomes the ground state of the system, see Figs.~\ref%
{profiles}(c) and (d). The orientation of the lattice is chosen randomly in
a process of spontaneous symmetry breaking, depending on the initial
fluctuations in the numerical minimization procedure. This symmetry breaking
is additional to the symmetry breaking associated with the randomly chosen
phase of the condensate wave function. In the reciprocal space, the
appearance of crystalline order is manifested by the appearance of side
peaks corresponding to the lattice wave vectors. We note that the lattice
constant of the pattern is determined by the wave vector value at the roton
minima. Indeed, the side peaks closest to $k=0$ in Fig.~\ref{profiles}(d)
match with the position of the roton minimum in Fig.~\ref{potentials}.

When decreasing the interlayer distance even further to $L=14\,$nm, the
minimization algorithm leads to the collapse of the wave function $\psi
_{\pm}$ into a single point of the numerical mesh. This indicates that the
ground state cannot be determined due to the breakdown of the mean-field
model. To obtain a correct ground state, one has to include higher order
effects that would limit the growth of the local condensate density. We note
that collapse is generally encountered for a nonlinear Schr\"{o}dinger
equation in the 2D (critical) case with attractive nonlinearity~\cite%
{Berge_collapse}.

In the supersolid regime ($L=18\,$nm), depending on the initial noise, the
minimization procedure can produce yet another type of solution, which
resembles a crystal structure with dislocations, see Fig.~\ref{profiles}(e).
While these solutions do not correspond to the ground state of the system,
they are local minima of the energy and constitute metastable states, and as
such are likely to appear in experimental realizations. In the momentum
space, these disordered supersolid patterns are characterized by concentric
rings around the $k=0$ point, see Fig.~\ref{profiles}(f).

\begin{figure}[tbp]
\includegraphics[width=8.5cm]{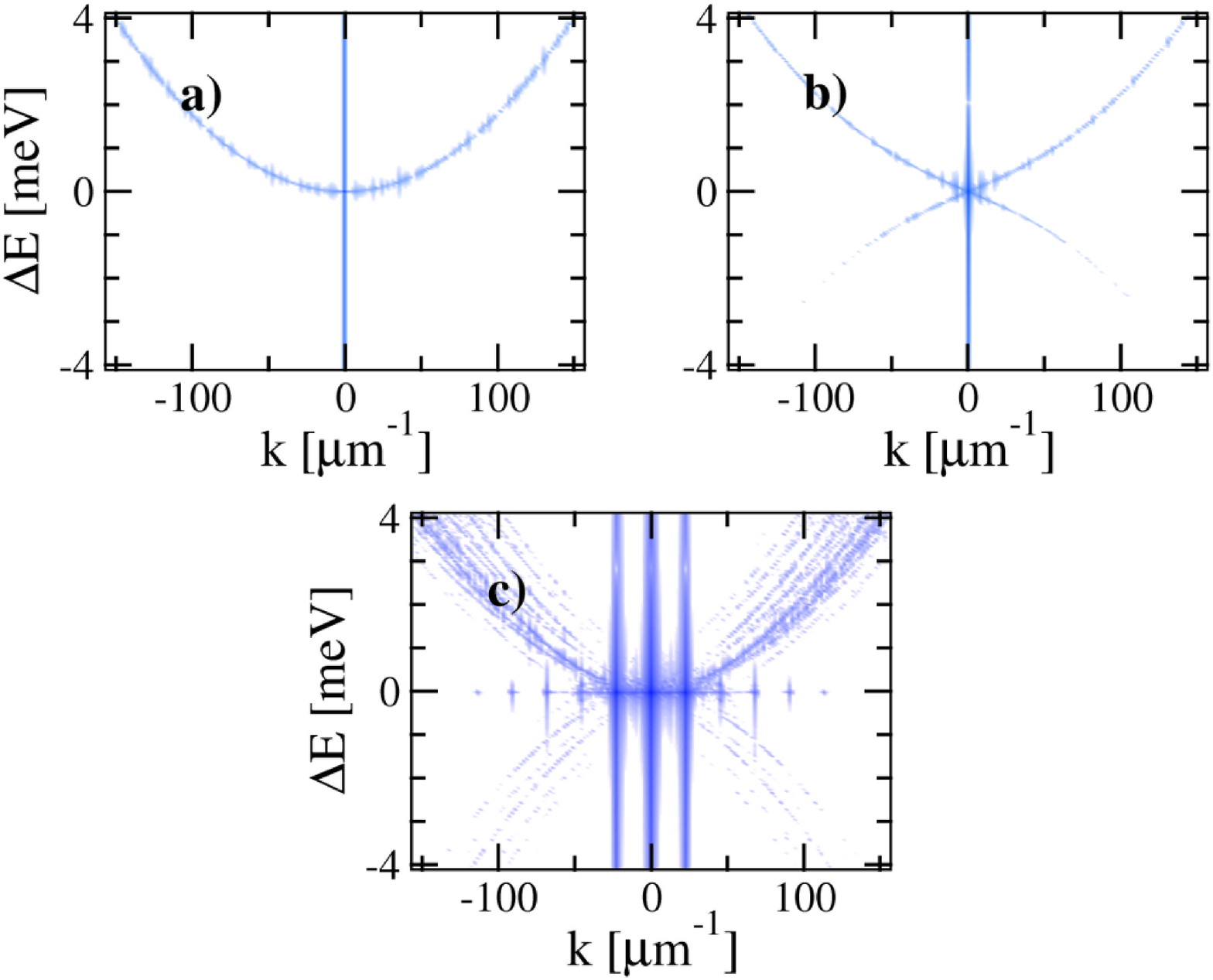}
\caption{Numerically determined spectra of a weakly perturbed (a)
noninteracting exciton condensate, (b) condensate interacting through a
local (momentum-independent) potential, and (c) supersolid condensate in the
presence of coupling with electron gas, interacting through an effective
potential $V^{\mathrm{eff}}_{\mathrm{ex-ex}}$ (parameters as in Figs.~%
\protect\ref{profiles}(c),(d)). While in the cases (a) and (b) we obtain the
parabolic and linear Bogoliubov spectra as predicted by the theory, in the
case (c) the spectrum is distorted as a result of interference of branches
starting from different $k$-components of the ground state in the momentum
space, corresponding to the vertical lines. The density is shown in a
logarithmic intensity scale. }
\label{spectra}
\end{figure}

We also investigate the influence of the supersolid transition on a weakly
excited system. In Fig.~\ref{spectra} we show the spectra of perturbed
ground states allowed to evolve in time according to Eq.~(\ref{GP}). We
obtain the spectra by Fourier transforming the wave function $\psi _{\mathrm{%
ex}}$ in space and time. The three figures correspond to the cases of a
noninteracting exciton condensate, a condensate with a $q$-independent
interaction potential, equivalent to a contact (local) repulsive interaction
in the absence of an electron gas, and an exciton supersolid interacting
through a $q$-dependent interaction potential $V_{\mathrm{ex-ex}}^{\mathrm{%
eff}}$. While the first two figures show the corresponding free-particle
parabolic spectrum and the Bogoliubov spectrum of excitations, respectively,
in the case of the supersolid the spectrum becomes blurred as a result of
interference of several branches starting from different points in the
momentum space. These points correspond to the $k$-space components of the
ground state, visible as pronounced vertical lines. The excitation branches
appear to be linear close to their points of origin, which is an indication
of a nonvanishing superfluid fraction.

\begin{figure}[tbp]
\includegraphics[width=6cm]{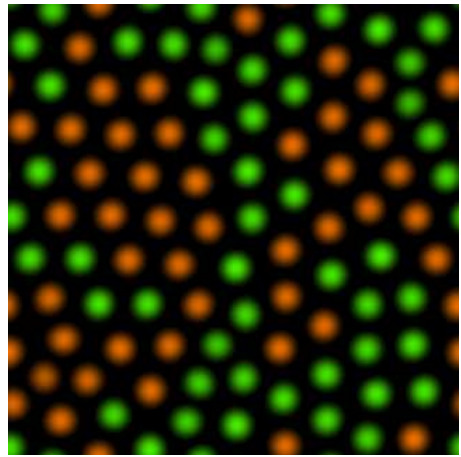}
\caption{ Numerically calculated density plot of the exciton condensate in
the supersolid phase, showing the ferromagnetic domain structure. Here $%
\overline{n}_{+}=\overline{n}_{-}= 10^{11} \mathrm{cm}^{-2}$, $L=18\,$nm and 
$\protect\alpha$ is chosen in such a way that for $q=0$ the interaction of
excitons with opposite spin is $10\%$ stronger than the interaction of
excitons with parallel spins. The green and orange coloring corresponds to
the $+1$ and $-1$ spin components. }
\label{ss_domains}
\end{figure}

Interesting new physics can be expected in two-component exciton
condensates, where $\alpha \neq 0$. Fig.~\ref{ss_domains} shows an example
of the spatial distribution of excitons in the supersolid phase in this
case. One can see that ferromagetic domains are formed: the triangular
crystal lattice is maintained in each domain, while long-range order is
broken by stochastic domain boundaries.

Experimentally, formation of the exciton supersolid may be detected by
angular and polarisation resolved photoluminescence, which would give access
to the characteristic diffraction pattern (most likely, concentric rings) in
the Fourier spectrum of excitations of the supersolid.

M.M. acknowledges support from the Foundation for Polish Science through the
\textquotedblleft Homing Plus\textquotedblright\ programme. TT and AK thank
Ivan Shelykh for useful discussions.


\begin{thebibliography}{99}
\bibitem{Supersolid_old} A. F. Andreev and I. M. Lifshitz, Sov. Phys. JETP 
\textbf{29} (1969) 1107; G.V. Chester, Phys. Rev. A \textbf{2}, 256 (1970);
A. J. Leggett, Phys. Rev. Lett. \textbf{25}, 1543 (1970).

\bibitem{Rotonlike} N. Henkel, R. Nath, and T. Pohl, Phys. Rev. Lett. 
\textbf{104}, 195302 (2010); F. Cinti \textit{et al.}, 
Phys. Rev. Lett. \textbf{105}, 135301 (2010); M. Matuszewski, Phys. Rev.
Lett. \textbf{105}, 020405 (2010).

\bibitem{Supersolidity_He} E. Kim and M.H.W. Chan, Nature (London) \textbf{%
427}, 225 (2004); E. Kim and M.H.W. Chan, Science \textbf{305}, 1941 (2004);
D. E. Galli and L. Reatto, J. Phys. Soc. Jpn. \textbf{77}, 111010 (2008); S.
Balibar, Nature (London) \textbf{464}, 176 (2010).

\bibitem{Lozovik2011} A. E. Golomedov, G. E. Astrakharchik, and Yu. E.
Lozovik, Phys. Rev. A \textbf{84}, 033615 (2011).

\bibitem{Butov} L.V. Butov, C.W. Lai, A.L. Ivanov, A.C. Gossard, and D.S.
Chemla,\emph{\ }Nature (London)\emph{\ }\textbf{417}, 47 (2002).

\bibitem{Kasprzak} J. Kasprzak \textit{et al.}, Nature (London), \textbf{443}%
, 409 (2006); R. Balili, V. Hartwell, D. Snoke, L. Pfeiffer and K. West,
Science, \textbf{316}, 1007 (2007).

\bibitem{Kavokin} A V Kavokin, J J Baumberg, G Malpuech, F P Laussy,
Microcavities, Oxford University Press (2007).

\bibitem{Amo} A. Amo \textit{et al.}, Nature (London) \textbf{457}, 291
(2009).

\bibitem{Ciuti} C. Ciuti \textit{et al.}, Phys. Rev. B \textbf{58}, 7926
(1998).

\bibitem{Rotons_Hybrid} I. A. Shelykh, T. Taylor, and A. V. Kavokin, Phys.
Rev. Lett. \textbf{105}, 140402 (2010).

\bibitem{Parish} M.M. Parish, F.M. Marchetti and P.B. Littlewood,
Europhysics Letters, \textbf{95, }27007 (2011).

\bibitem{Lozovik} Yu. E. Lozovik and V. I. Yudson, Pis'ma Zh. Eksp.
Teor.Fiz. \textbf{22}, 556 (1975) [JETP Lett. \textbf{22}, 274 (1975)]; R.
Zimmermann, Phys. Status Solidi B 243, 2358 (2006); Z. V\"{o}r\"{o}s, D. W.
Snoke, L. Pfeiffer, and K. West, Phys. Rev. Lett. \textbf{103}, 016403
(2009).

\bibitem{Yudson} V.I. Yudson, Phys. Rev. Lett. \textbf{77}, 1564 (1996).

\bibitem{Laussy} F.P. Laussy, A.V. Kavokin and I.A. Shelykh, Phys. Rev.
Lett. \textbf{104}, 106402 (2010).

\bibitem{Combescot} M. Combescot, O. Betbeder-Matibet, and R. Combescot,
Phys. Rev. Lett. \textbf{99}, 176403 (2007).

\bibitem{Sinclaire} N. W. Sinclair \textit{et al.}, Phys. Rev. B \textbf{83}%
, 245304 (2011).

\bibitem{Leggett_BEC} A. J. Leggett, Rev. Mod. Phys. \textbf{73}, 307 (2001).

\bibitem{Laikhtman} B. Laikhtman and R. Rapaport, Phys. Rev. B \textbf{80},
195313 (2009).

\bibitem{Zimmermann} C. Schindler and R. Zimmermann, Phys. Rev. B \textbf{78}%
, 045313 (2008).

\bibitem{Ramon} G. Ramon \textit{et al.}, Phys. Rev. B \textbf{65}, 085323
(2002).

\bibitem{Berge_collapse} L. Berg\'{e}, Physics Reports \textbf{303}, 259
(1998).
\end{thebibliography}
\end{document}